# Versatile microwave-driven trapped ion spin system for quantum information processing[a]

Christian Piltz[1], Theeraphot Sriarunothai[1], Svetoslav S. Ivanov[2], Sabine Wölk[1], Christof Wunderlich[1]*

[1]Department Physik, Naturwissenschaftlich-Technische Fakultät, Universität Siegen, 57068 Siegen, Germany.
[2]Department of Physics, Sofia University, 5 James Bourchier Boulevard, 1164 Sofia, Bulgaria.
*Corresponding author. E-mail: christof.wunderlich@uni-siegen.de

**Abstract**
Using trapped atomic ions we demonstrate a tailored and versatile effective spin-system suitable for quantum simulations and universal quantum computation. By simply applying microwave pulses, selected spins can be decoupled from the remaining system and thus can serve as a quantum memory, while simultaneously, other coupled spins perform conditional quantum dynamics. Also, microwave pulses can change the sign of spin-spin couplings, as well as their effective strength, even during the course of a quantum algorithm. Taking advantage of the simultaneous long-range coupling between three spins a coherent quantum Fourier transform – an essential building block for many quantum algorithms – is efficiently realized. This approach, which is based on microwave-driven trapped ions and is complementary to laser-based methods, opens a new route to overcoming technical and physical challenges in the quest for a quantum simulator and a quantum computer.

**INTRODUCTION**

The idea of using one quantum system under sufficient experimental control to investigate the static and dynamic properties of another, complex quantum system that is difficult to access directly (*1*) has spurred the development of proposals for universal quantum computing and for quantum simulations. Devices for processing quantum information promise new insight into pertinent scientific problems ranging from quantum field theories, to molecular dynamics, to cosmology to name but a small sample (*2-4*). A universal quantum computer would be suitable for tackling any of these scientific questions. More specialized quantum simulations – already in close experimental reach – also promise groundbreaking new insight into phenomena governed by quantum processes.

Recently, important experimental progress was demonstrated with several auspicious, potentially scalable physical systems, including trapped neutral atoms (*5*), propagating photons (*6*), superconducting qubits (*7*), and spins in semiconductors (*8*). Trapped atomic ions offer a level of quantum control of individual quantum systems that is unsurpassed, and systems of coupled effective spins realized by trapped ions are a powerful resource not only for universal quantum computing (*9-11*) but also for specific quantum simulations (*12-19*).

---

[a] Journal Reference: Science Advances **2,** e1600093 (2016)

Here, we explore a system of effective spins represented by individual trapped ions that exhibit long-range spin-spin coupling. Such a long-range coupling between effective spins can be induced by a spatially varying static or dynamic magnetic field leading to a state dependent shift of the ions' internal states (*14,20-24*), or by laser beams inducing a state dependent optical dipole force (*10-19,25*).

In the case of magnetically induced coupling, coherent operations are carried out using radio-frequency (RF) radiation, and the application of global pulses is sufficient to control coherent interactions between spins thus avoiding technical and physical challenges associated with the use of laser light (*26-31*).

Quantum simulations carried out on a physical system that is particularly well suited for solving a particular problem are likely to yield new insight that is not available from computations on classical computers. A universal quantum computer could perform any such quantum simulation and, in addition, solve other problems that are intractable, for all practical purposes, on classical computers (*32-36*). Various implementations of essential elements of such a device and even complete quantum algorithms have been carried out (*37*), with atomic trapped ions playing a prominent role.

Here, we take advantage of magnetic gradient induced coupling (MAGIC) between effective spins of trapped ions. We experimentally realize different coupling topologies within a three-spin-1/2 system, and show ferromagnetic and antiferromagnetic interactions. All coupling topologies are attained by individual ion control using microwave (MW) pulses (*38*). Hence, during a quantum simulation or quantum algorithm, one can easily and rapidly change ("on the fly") from one interaction to another. Furthermore, we also decouple any desired spin from the rest of the system. This enables one to use a subset of the qubit register as a quantum memory while simultaneously carrying out conditional quantum gates with other qubits. In addition, the change between coupling topologies enables the efficient implementation of quantum algorithms that are useful for quantum information processing, which is demonstrated by realizing a coherent quantum Fourier transform (QFT), taking advantage of simultaneous coupling between three spins.

**RESULTS**
**Experimental setup**
$^{171}$Yb$^+$ ions are confined in a linear Paul trap with an axial trap frequency of $\nu_1 = 2\pi \times 130\,\text{kHz}$ and a radial trap frequency of $2\pi \times 500$ kHz. After Doppler cooling the ions form a linear Coulomb crystal (*39*). Two hyperfine states of each ion's ground state $^2S_{1/2}$ represent an effective spin 1/2. If this spin is formed using at least one magnetically sensitive state, and in the presence of a static magnetic field gradient, effective spins can be distinguished by their frequency for resonant excitation. Also, they are coupled via a pairwise spin-spin interaction (*14,20-22,29*). If the spin up state is encoded in the magnetic sensitive level $\left|^2S_{1/2}, F=1, m_F = \pm 1\right\rangle \equiv \left|\uparrow\right\rangle$ the (linear) Zeeman effect causes the internal energy of spin $i$ to depend on the ion position as $\hbar\omega_i = g_F m_F \mu_B B(z_i)$, where $g_F$ denotes the Landé factor, $m_F$ denotes the magnetic quantum number (±1 for magnetic $\sigma^\pm$ transitions used here), $\mu_B$ denotes the Bohr magneton and $B(z_i)$ denotes the magnetic field at qubit $i$'s axial position $z_i$.

The magnetic gradient present in our setup is 19 T/m (*29*). An additional magnetic bias field and the trapping potential described above result in magnetic fields at the ions' positions of 0.1881(4), 0.4146(4), and 0.6432(5) mT. The corresponding addressing frequency separations are about 3.2 MHz. Hence each of these spins can be individually excited by dialing in the MW frequency near 12.6 GHz matching the respective spin resonance (*38*).

The Hamiltonian describing the spin-spin interaction of $N$ ions reads (*21,22,29*):

$$H_\mathrm{I} = -\frac{\mathrm{h}}{2} \sum_{\substack{i,j=1 \\ i \neq j}}^{N} J_{ij} \sigma_z^{(i)} \sigma_z^{(j)}, \qquad (1)$$

where h denotes the Planck constant and $\sigma_z^{(i)}$ denotes the z Pauli matrix on qubit $i$'s subspace.

The mutual coupling strengths are given by

$$J_{ij} = \sum_{n=1}^{N} \nu_n \epsilon_{in} \epsilon_{jn} \qquad (2)$$

with the dimensionless constants $\epsilon_{in} \equiv \partial_z \omega_i \frac{\Delta z_n}{\nu_n} S_{in}$. The dimensionless entry $S_{in}$ of the unitary matrix that diagonalizes the dynamical matrix of the system is the scaled deviation of ion $i$ from its equilibrium position when vibrational mode $n$ is excited. Hence, $\epsilon_{in}$ describes how strongly the spin of this ion couples to the vibrational mode. Here, $\Delta z_n = \sqrt{\mathrm{h}/2m\nu_n}$ denotes the extension of the ground state wave function of vibrational mode $n$, which has the angular frequency $\nu_n$, and $m$ is the ion's mass.

Because the partial derivative of the internal energy of the spins always occurs pairwise in the J-coupling strength between two spins [$(\partial_z \omega_i) \times (\partial_z \omega_j)$ in Eq. (2)], it is possible to control the sign of their mutual coupling. If the two spins have their spin up state encoded in different hyperfine levels $|{}^2S_{1/2}, F=1, m_F = \pm 1\rangle$, then their magnetic quantum numbers $m_F$ differ in sign and, thus, their mutual J-coupling will be negative. However, if the magnetic quantum numbers are equal the J-coupling will be positive.

When choosing (to first order) magnetically insensitive spin states, $|{}^2S_{1/2}, F=0\rangle \equiv |\downarrow\rangle$ and $|{}^2S_{1/2}, F=1, m_F = 0\rangle \equiv |\uparrow\rangle$ (referred to as the $\pi$ basis in what follows) all of the spins have the same resonance frequency of the magnetic $\pi$ transition; in addition, spin-spin coupling is suppressed.

**Fully coupled three-spin systems**

To experimentally determine the joint couplings within a three-spin system, we observe its internal dynamics for different initial states. All spins are encoded in the hyperfine levels $|\downarrow\rangle \equiv |{}^2S_{1/2}, F=0\rangle$ and $|\uparrow\rangle \equiv |{}^2S_{1/2}, F=1, m_F = -1\rangle$ (henceforth referred to as the $\sigma^-$ basis). The dynamics are studied using Ramsey-like experiments (see Materials and Methods), where one of the spins is prepared in a superposition state, whereas the other two are prepared in energy eigenstates. During a conditional evolution time the system is left to evolve according to the time evolution $U(T) = e^{-\frac{i}{\mathrm{h}} H_\mathrm{I} T}$ which results in a precession of the spin in the superposition state, whereas the spins in energy eigenstates are left unaffected. The spin precession is revealed by observing Ramsey interference fringes. Because the conditional dynamics are slower than the inverse dephasing time of a single spin we apply 20 dynamical decoupling (DD) pulses during the conditional evolution time to protect the dynamics (see Materials and Methods) (*40*).

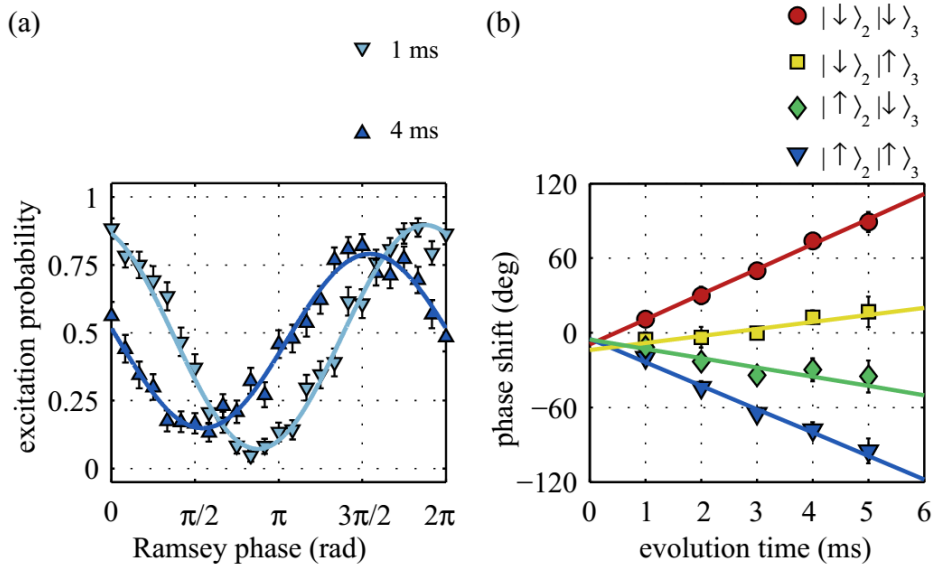

**Fig. 1: Conditional quantum dynamics in a fully coupled three-spin system.** (a) Ramsey fringes from the first spin after different conditional evolution times (1 and 4 ms). The phase shift reveals the coupling to other spins. Each data point represents the result of 50 repetitions of the experiment. (b) For different initial states of the second and third spin and different conditional evolution times the acquired phase shift is plotted. From the slope, information about the couplings can be deduced. Error bars represent standard deviations.

Exemplary results indicating conditional precession of spin 1 are shown in Figure 1. The resulting Ramsey fringes with spins 2 and 3 prepared in $\left|\uparrow\right\rangle_2\left|\uparrow\right\rangle_3$ (the indices denote the spin) are shown in Figure 1A for two different conditional evolution times (1 and 4 ms). As one can see the fringe minimum, which is found at $\pi$ for vanishing conditional evolution time, is shifted to smaller phases for increasing conditional evolution time which reveals the spin precession. One may also notice a reduction of the fringe contrast which can be explained mainly by dephasing because of magnetic field noise that is not fully compensated by the present DD sequence. We measure the dephasing time of a single spin as $\tau^\sigma = 7.1(1.1)$ ms when 20 DD pulses are applied to each spin (see Materials and Methods).

This experiment is repeated for different input states of the second and third spins and for different conditional evolution times. In Figure 1B the acquired phase shifts conditioned on the initial states of spins 2 and 3 are shown. From these phase shifts $\Delta\varphi = JT$, acquired by spin 1 during a conditional evolution time $T$, we obtain information about the couplings among the system. For the symmetric input state $\left|\uparrow\right\rangle_2\left|\uparrow\right\rangle_3$ the time evolution of the first spin is described by $e^{-i(J_{12}+J_{13})\sigma_z^{(1)}t/2}$ which is a precession around the $z$-axis at angular velocity $J = J_{12} + J_{13}$. Hence, from the observed phase shifts we conclude that $J_{12} + J_{13} = 2\pi \times 52(9)$ Hz. For the other symmetric state $\left|\uparrow\right\rangle_2\left|\uparrow\right\rangle_3$ the evolution corresponds to a precession about the negative $z$ axis at the same angular velocity $J_{12} + J_{13}$. For the antisymmetric input state $\left|\uparrow\right\rangle_2\left|\uparrow\right\rangle_3$ the time evolution is described by $e^{-i(J_{12}-J_{13})\sigma_z^{(1)}t/2}$ and from the phase shifts we conclude that $J_{12}-J_{13}=2\pi\times 21(9)$ Hz.

The other antisymmetric input state $|\uparrow\rangle_2|\uparrow\rangle_3$ causes a spin precession in the other direction at the same angular velocity. Here, the value of the precession frequency derived from a fit is $2\pi\times 16(9)$ Hz. For the symmetric input states the angular velocity is larger; hence, the slopes extracted from a fit of the data are less affected by statistical errors.

To measure the mutual coupling between two spins among the three-spin system, we effectively suppress the third spin's coupling by selective recoupling (see Materials and Methods) (*41*). Here, an additional spin echo (SE) $\pi$ pulse is addressed to the third spin and cancels the phase shift due to this spin. From this we reconstruct the J-coupling matrix shown in Figure 2A with all couplings being positive.

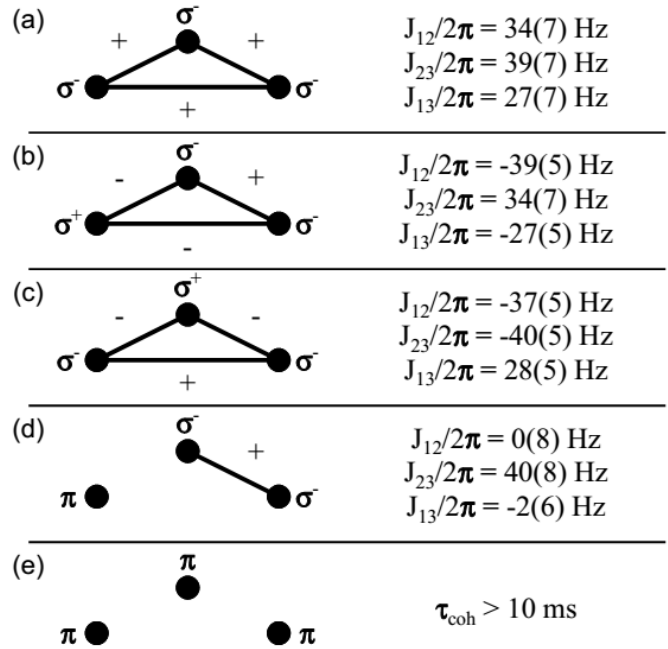

**Fig. 2: Different experimentally realized spin 1/2 systems.** Spins are symbolized by dots and the base in which each spin is encoded is denoted by $\sigma^{\pm}$ and $\pi$, respectively. A line connecting two dots represents a J-coupling between them, whereas $\pm$ indicates the sign of this interaction. Experimental results are summarized in the right column. (a) All spins are encoded in the same basis; hence, their coupling has a positive sign. (b) and (c) One of the spins is encoded in a different base, which results in different interactions. (d) It is also possible to decouple a selected spin while the remaining spins are left coupled. (e) If the coupling of all spins is suppressed their states are preserved.

**Changing the coupling topology**

We realize spin systems with different signs of the couplings by coding the spin states in different hyperfine levels. In contrast to the experiments described above, one of the spin up states is now encoded in $|\uparrow\rangle\equiv|^2S_{1/2},F=1,m_F=+1\rangle$ and this spin is realized by the magnetic $\sigma^+$ transition of its ion. The J-coupling matrices of such systems are reconstructed, and we summarize the results in Figure 2 (B and C). The topology from Fig. 2B features next-neighbor couplings of opposite signs, whereas in the topology from Fig. 2C the next-neighbor couplings have the same sign.

As we have shown, single-ion control allows for preparing any spin in different basis states thereby tailoring the couplings in the system. In addition, it is possible to change the spin states during a quantum algorithm by application of MW pulses, which will change the couplings accordingly. Therefore, the change between different couplings can be performed on the time scale of the Rabi frequency $\Omega$, that is, nearly instantaneously on the time scale of the conditional evolution. In these experiments, $\Omega \approx 2\pi \times 50$ kHz.

We demonstrate this change between two bases by preparing the first spin in a superposition state in the $\sigma^-$ basis. Then, we transfer the state by a sequence of MW pulses (see Materials and Methods) to the $\pi$ basis where it is decoupled from the rest of the spin system. After a conditional evolution time, the spin state is transferred back to the $\sigma^-$ basis and the coherence is probed with a Ramsey pulse as before. Now, this spin does not acquire any phase shift that depends on the other spins' initial states. Because that this spin is less affected by magnetic field fluctuations we measure its coherence time to be $\tau^\pi = 50(10)$ ms when only a single DD pulse is applied. The coupling between spins 2 and 3 is, again, reconstructed by Ramsey-like experiments and experimental results are shown in Figure 2D. In the context of quantum information processing such a system of two coupled spins and one uncoupled spin with a long coherence time realizes a quantum memory, while simultaneously the other two spins may perform a conditional quantum gate.

Furthermore, it is possible to simultaneously suppress all couplings within the system by transferring all the spin states to the magnetic insensitive base (compare Figure 2E). Experimentally we protect the product state of equal superposition states $1/\sqrt{8}\left(\left|\downarrow\right\rangle+i\left|\uparrow\right\rangle\right)\left(\left|\downarrow\right\rangle+i\left|\uparrow\right\rangle\right)\left(\left|\downarrow\right\rangle+i\left|\uparrow\right\rangle\right)$ (hereafter, indices are omitted for simplicity) for 10 ms and detect a preserved coherence, for instance, of the second spin having a fidelity of $0.91(2)$.

**Quantum Fourier transform**

Fully coupled spin systems are a resource for quantum computation, where each spin represents a qubit. We present the realization of an example quantum algorithm: the QFT, defined with its action on an orthonormal basis $\left|0\right\rangle, \left|1\right\rangle,\ldots, \left|N-1\right\rangle$,

$$F^N \left|n\right\rangle = \frac{1}{\sqrt{N}} \sum_{k=0}^{N-1} e^{2\pi i n k/N} \left|k\right\rangle. \qquad (3)$$

We have chosen this particular algorithm because of its relevance in providing exponential speedup as a subroutine in many quantum algorithms (*35,42,43*). Previous realizations of the QFT in systems based on nuclear magnetic resonance (*44,45*) rely on single-qubit gates and conditional two-qubit gates. A realization with trapped ions was achieved using single qubit gates, collective nonentangling gates and collective entangling gates (*46*). A semiclassical QFT, without conditional two-qubit gates, using trapped ions is described by Chiaverini *et al.* (*47*). Such a realization can be used for example in the Shor algorithm (*32*) where the QFT is the last subroutine before a projective measurement of the output state. Because only probability amplitudes (and no relative phase information) need to be deduced, the fully coherent QFT can be replaced by the semiclassical version in such cases. However, there are existing also algorithms in which the QFT or the inverse QFT is not the final subroutine and phase information of the output states is of further importance [for example, Harrow *et al.* (34) and Wiebe *et al.* (36)]. In these algorithms the QFT cannot be replaced by a semiclassical version; hence, the fully coherent QFT is necessary for their realization.

In physical systems that are described by an Ising-type Hamiltonian, usual realizations of the QFT demand for the suppression of undesired couplings among the system during the conditional evolution of two selected qubits. However, in general, it is favorable to decompose the QFT into gates that can be realized naturally in the Ising-type system, for example, single-qubit operations and waiting times, during which the system evolves according to all the mutual couplings (*48-51*). In this way, the available physical system is efficiently used to achieve a reduced operation time and, consequently, to improve the fidelity of the quantum process, if the physical system is subjected to decoherence.

In what follows, we describe our gate set, which comprises a rotation $R(\theta)$, a phase gate $\Phi(\phi)$, and an entangling gate $U(T)$. We will also use phased rotations $R(\theta,\phi)$, where the phase gate is absorbed in the rotation according to the rule $R(\theta,\phi) \equiv \Phi(\phi/2)R(\theta)\Phi(-\phi/2)$. The gate $R(\theta,\phi)$ can be obtained with a simple phase shift in the driving MW field implementing $R(\theta)$. We have

$$R_k(\theta,\phi) = e^{-i\frac{\theta}{2}(\sigma_x^{(k)}\cos\phi + \sigma_y^{(k)}\sin\phi)}, \quad (4)$$

$$\Phi_k(\phi) = e^{-i\phi\sigma_z^{(k)}}, \quad (5)$$

where the indices denote the qubit number and $R_k(\theta) = R_k(\theta,0)$.

The entangling gate is easily obtained from free evolution: We leave the ions to evolve with no driving for a certain conditional evolution time $T$,

$$U_{kl}(T) = e^{iJ_{kl}T/2\sigma_z^{(k)}\sigma_z^{(l)}}, \quad (6)$$

where $J_{kl}$ represents the coupling strength for qubits $k$ and $l$.

We derive the following sequence, which implements the QFT (up to a global phase) based on single-qubit rotations and entangling gates:

$$\begin{aligned} U_{\mathrm{QFT}} = &\, R_2\left(A_2, \tfrac{3\pi}{4}\right) R_3\left(\tfrac{\pi}{2}, -\tfrac{\pi}{2}\right) U_{23}(T_3) R_1\left(\pi, \tfrac{3\pi}{16}\right) R_2\left(A_1, \tfrac{3\pi}{4}\right) \\ &\, R_3\left(\pi, -\tfrac{3\pi}{16}\right) U(T_2) R_3(\pi) U(T_1) R_1(\pi) R_2(\pi) R_3(\pi) H_1. \end{aligned} \quad (7)$$

The derivation is described in more detail by Ivanov *et al.* (*52*) and in Materials and Methods where the duration $T_3$ and the pulse areas $A_1$ and $A_2$ are also discussed. For the typical experimental parameters of the setup (*29*) one obtains the evolution times $T_1$=3.69 ms, $T_2$=0.22 ms, $T_3$=4.87 ms. Hence, the total duration of the algorithm is 8.78 ms. For the pulse areas, one obtains $A_1 = 0.686\pi$ and $A_2 = 0.716\pi$.

The sequence in Eq. (7) can be further optimized. We first note that $T_2$ is much shorter than the other conditional evolution times, and can therefore be neglected which results in an infidelity on the order of $10^{-3}$. By making use of the fact that the entangling gate and simultaneous $\pi$ pulses applied to all three qubits commute, we combine subsequent pulses applied to a qubit whenever possible. The gate $U_{23}(T_3)$ where the first qubit's coupling is suppressed could be realized by a change of the qubit base, as shown in our experiments. However, for simplicity it is realized here by an SE pulse in the middle of the evolution time. An additional $\pi$ pulse compensates for the state change of the first qubit and takes into account a phase gate that originates from the combination of subsequent $\pi$ pulses. The optimized sequence reads

$$U_{QFT}^{opt} = \left[ R_2(A_2, \frac{3\pi}{4}) R_3(\frac{\pi}{2}, \frac{3\pi}{2}) R_1(\pi, \frac{27\pi}{16}) \right] \left[ U(T_3/2) R_1(\pi, \frac{\pi}{2}) U(T_3/2) \right]$$
$$\left[ R_2(A_1, \frac{3\pi}{4}) R_2(\pi, 0) R_3(\pi, \frac{13\pi}{16}) \right] U(T_1) H_1. \tag{8}$$

and the circuit is presented in Figure 3. An additional SWAP$_{13}$ gate that interchanges the state of qubits 1 and 3 completes the QFT. This gate is realized by just relabeling the qubits.

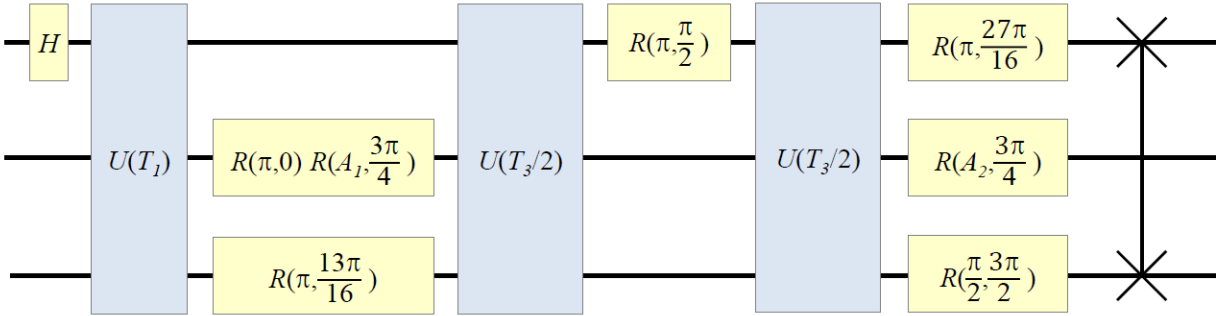

**Fig. 3: Circuit for the QFT.** The circuit that realizes the QFT in our experiments consists of single-qubit rotations [$R(\theta, \phi)$] and conditional evolutions [$U(T)$] of different duration. For technical reasons single-qubit operations on different qubits are applied one after another and the Hadamard gate ($H$) is realized by two rotations [$R(\pi/2, -\pi/2) R(\pi, 0)$]. The additional SWAP gate that completes the QFT is realized by relabeling qubits 1 and 3 after the projective measurement. During the optimization of the sequence the conditional evolution time $T_2 = 0.22$ ms was neglected. The whole sequence takes 8.6 ms as a result of the conditional evolution times. The duration of the single-qubit rotations can be neglected because the Rabi frequency is about $2\pi \times 50$ kHz.

The chosen realization takes advantage of the fully coupled three-qubit system and consists only of conditional free evolution times and single-qubit gates (see Materials and Methods) (*52*). In comparison to the serial decomposition of the QFT using two-qubit gates, which would take about 23 ms in our setup, the duration of this efficient sequence is only 8.6 ms. The relative advantage in terms of time grows further with growing $N$ (*50*).

A first validation of the desired quantum dynamics is achieved by simultaneous Ramsey-type measurements after the three-qubit register was prepared in one of the eight computational basis states $|000\rangle, |001\rangle, |010\rangle, |011\rangle, |100\rangle, |101\rangle, |110\rangle, |111\rangle$. Because the QFT transfers each computational basis state to a product state of superpositions, we apply to each individual qubit a Ramsey $\pi/2$ pulse to probe the qubit states. In Figure 4 exemplary results for the input state $|010\rangle$ are shown. The solid lines are fits to the data points, whereas the light dashed lines represent the ideal outcome of the QFT. The Ramsey fringes' phases match the expected result of a QFT, which is also true for all the other computational basis states chosen as initial states. The average deviation between the expected and the experimentally obtained phase is $\langle \delta\varphi \rangle = -0.07(15)$ rad. The phase of each qubit is determined by all couplings present in the three-qubit system. Therefore, we conclude that the conditional evolution times and single-qubit gates chosen to implement the QFT match the couplings among the system.

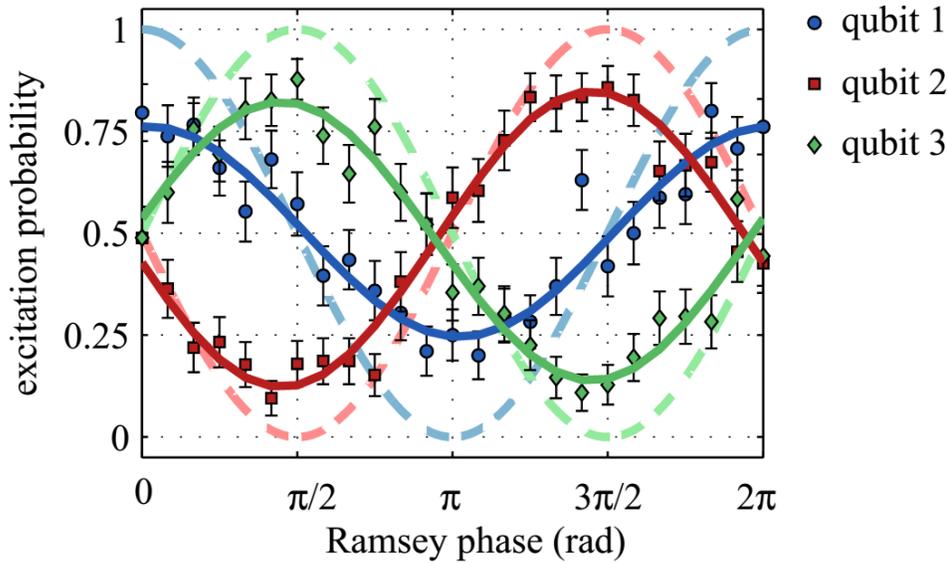

**Fig. 4: Exemplary results of a QFT.** After the QFT with input state $|010\rangle$, additional Ramsey $\pi/2$ pulses reveal interference fringes from every single qubit. The experimental results (solid lines) are compared with the predicted outcomes (dashed lines). Each data point represents 50 repetitions and error bars denote standard deviations.

Although the phases of the fringes match the expectations, their contrast is significantly reduced. We explain this by dephasing during the conditional evolution periods despite applying DD, by pulse imperfections of the 60 DD pulses applied during the complete QFT, and by a limited single-shot readout fidelity (see Materials and Methods).

The three single qubit output fidelities for input state $|010\rangle$ are $F_1 = 0.76(7)$, $F_2 = 0.86(4)$, and $F_3 = 0.84(4)$. If the output state is a fully separable state, then the three-qubit state fidelity is simply given by $F_1 \times F_2 \times F_3$. For a general state, the three-qubit state fidelity $F$ of the state $\rho$ right after application of the QFT is given by $F = \langle \psi | \rho | \psi \rangle$, where $|\psi\rangle$ is the ideal outcome. We further quantify the performance of the QFT by reconstructing this fidelity and obtain an average fidelity $\langle F \rangle = 0.58(5)$. Fidelities of the QFT for all input states (the basis states mentioned above and, in addition, for several superposition states) are given in Materials and Methods. Because the QFTs of the computational basis states are given by superposition states with equal weights for each qubit, an average fidelity of $\langle F \rangle = 0.58(5)$ clearly indicates nonclassicality. For comparison, an average fidelity of $0.5^3 = 0.125$ can be achieved with classical states.

In addition to validating the process that realizes the QFT for computational basis states and some exemplary superposition states, we demonstrate that the algorithm correctly estimates the period of quantum states. After initializing the register in the desired input state of interest the sequence is applied and a projective measurement takes place. The probability of each possible outcome ($|000\rangle, |001\rangle, ..., |111\rangle$) is recorded. In binary notation these states represent the integer numbers $i = 0, 1, 2 ... 7$ which label the output states. The necessary $SWAP_{13}$ operation that completes the QFT is realized by interchanging qubits 1 and 3 during the analysis.

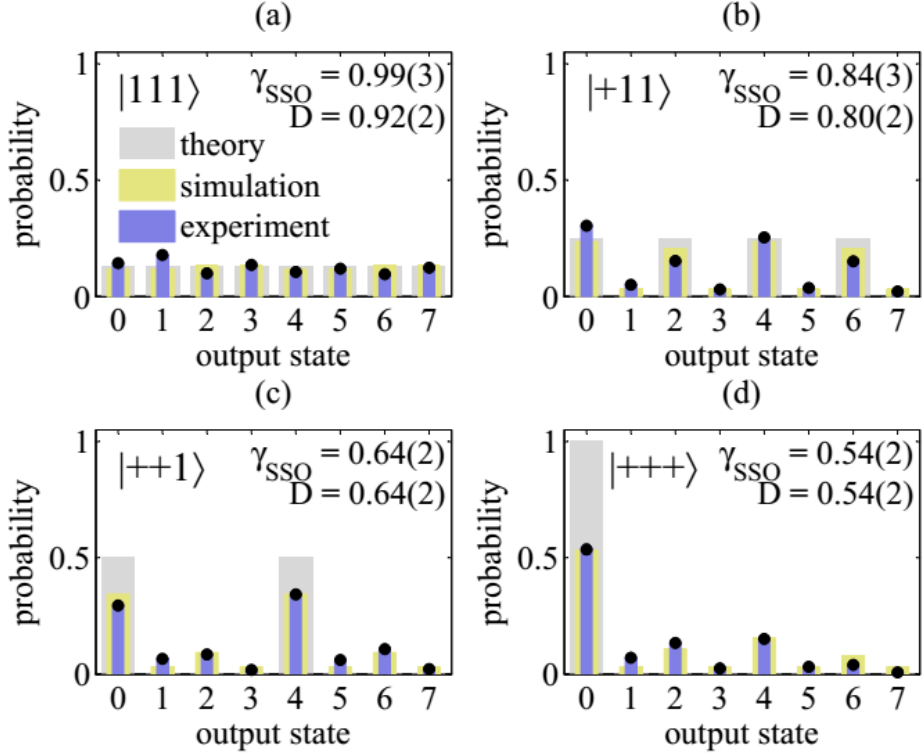

**Fig. 5: Estimating the period of quantum states.** (a) to (d) After application of the QFT for different input states, a projective measurement takes place. The measured probability of finding each possible output state is shown (blue bars) along with the ideal results (gray bars) and with simulations that take into account experimental imperfections of the setup (yellow bars). The SSO $S$ and distinguishability $D$ are measures of the performance. Each experiment is repeated 1250 times. The statistical error is too small to be shown.

Figure 5 shows the results for the four input states $|111\rangle, |+11\rangle, |++1\rangle$ and $|+++\rangle$ which are of periods 8, 4, 2 and 1 respectively. We compare the experimental results both to the ideal expectation and, in addition, to simulations that take into account the experimental imperfections present in the setup. Suitable measures that allow for a comparison of the experimentally determined probability $p_i$ with the expected correlations $q_i$ are the squared statistical overlap (SSO) $\gamma_{SSO}(p,q) = (\sum_i \sqrt{p_i \times q_i})^2$ and the distinguishability $D(p,q) = 1 - 1/2 \sum_i |p_i - q_i|$ (53).

As one can see the SSO and the distinguishability for the different states vary from 0.99(3) to 0.54(2) for the input states $|111\rangle$ and $|+++\rangle$, respectively. The reason for this broad range is the different susceptibility of different states to technical imperfections. One further observes that the match of the classical measures of the SSO and the distinguishability can be higher than the mean quantum state fidelities of 0.58(5) that have been discussed above. This can be explained by the fact that these benchmarks, in contrast to the quantum state fidelities, do not measure quantum phase relations.

## DISCUSSION

MAGIC makes it possible to coherently manipulate trapped ions using exclusively MW and RF radiation for all coherent operations. In similar laser-based experiments with trapped ions, signals controlling the coherent dynamics of ions are generated first in the MW or RF regime; then laser light is modulated using these signals. Here, we avoid this detour via the optical regime, demanding the precise control of coherent light, and thus considerably simplify experimental requirements. In addition, spontaneous scattering is practically eliminated and the vulnerability of quantum logic operations to motional excitation is reduced. The latter is true as long as the ions experience a harmonic potential which is the case for Doppler cooled ions as has been investigated in much detail theoretically and in recent experiments (to be published). Also, individual addressing of interacting spins is possible with unprecedented low cross-talk (*38*).

We show that the spin-spin coupling topology can be experimentally tailored to best suit a desired quantum simulation or computation, and we have investigated different coupling topologies among an exemplary three-spin system realized by single-ion control using MW pulses. We demonstrate how to rapidly change the sign of the couplings in the system. Furthermore, any spin can be decoupled from the remaining coupled system. Thus, switching between the role of an ion as a memory qubit on the one hand, and a qubit conditionally processing quantum information on the other hand, is realized by simple MW pulses and does not require physical relocation (shuttling) of ions, or the use of different ion species.

Also, the effective magnitude of spin-spin coupling during a quantum simulation or computation can be adjusted by varying the interaction time. In addition, the coupling strength can be varied by adjusting the trapping potential (*29*) and the magnetic gradient.

In microstructured segmented ion traps (*54*) detailed control of (quasi-)static DC potentials is possible. These potentials determine local and global trapping potentials, thus allowing for control of the range of interaction (for example, from long range to next-neighbor). At the same time, larger magnetic gradients can be achieved in microstructured traps. Thus, the use of such traps adds more versatility for realizing coupling topologies suitable for quantum simulation and computation (*55-59*).

A QFT that may serve as an essential ingredient of other quantum algorithms is implemented here. We take advantage of simultaneous coupling between all qubits, and a speedup of nearly a factor of 3 is achieved for three qubits as compared with a decomposition of the QFT into two-qubit gates.

The spin-spin coupling used here increases quadratically with larger magnetic gradients. In recent experiments a gradient of 150 T/m (*60*) has been implemented which would already yield an increase in coupling strength by about a factor of 60 as compared to the experiments reported here. In this case, the entangling gate time of the algorithm is reduced by the same factor to about 140 µs, and DD is no longer required. Single-qubit MW gates as used here have been previously realized with outstanding fidelity (*31,61*). Thus, a fidelity of the QFT higher than 0.99 is realistic when using a larger magnetic gradient (see Materials and Methods).

Because of the generality of the method used here, which takes advantage of a fully coupled spin system, this approach is attractive for other physical systems as well. A three-qubit processor could be used as an elementary unit for distributed quantum computing (*66*). Also, a relatively small quantum processor with qubits that serve at the same time as a quantum memory and a quantum processor for simultaneous conditional quantum dynamics is well suited as an essential element of a quantum repeater (*67*).

Using a larger magnetic gradient would also be useful for quantum simulations of various spin systems with flexible control of the coupling topologies. Furthermore, the system and methods described here are well suited for other quantum simulations, for instance, of lattice gauge theories (*68*).

## MATERIALS AND METHODS

### Spin precession and Ramsey experiments

We used Ramsey-like experiments to observe the ions' internal coherent dynamics and deduce the couplings among the spins from these measurements. In this experiment, all of the spins were prepared first in $|\downarrow\downarrow\downarrow\rangle$. Then, one of the spins was prepared in a superposition state $|\psi\rangle = \frac{1}{\sqrt{2}}(|\downarrow\rangle + i|\uparrow\rangle)$. The other two spins were prepared in one of the energy eigenstates $\{|\downarrow\downarrow\rangle, |\downarrow\uparrow\rangle, |\uparrow\downarrow\rangle, |\uparrow\uparrow\rangle\}$. During a conditional evolution time $T$ the system evolved according to

$$U(T) = e^{i(J_{12}\sigma_z^{(1)}\sigma_z^{(2)} + J_{23}\sigma_z^{(2)}\sigma_z^{(3)} + J_{13}\sigma_z^{(1)}\sigma_z^{(3)})T/2}. \quad (9)$$

Hence, the spins in energy eigenstates will not change their states, whereas, the spin in the superposition state will perform a spin precession conditioned on the state of the other two spins. After the conditional evolution time $T$ a Ramsey $\pi/2$ pulse was addressed to the spin in the superposition state and the spin's state was measured. The phase of this pulse was varied while repeating the experiment in order to record Ramsey interference fringes. The phase information of these fringes reveals information about the spin state. Therefore, by observing the Ramsey fringes changing their phase for a varying conditional evolution time we deduced the angular frequency of the spin precession.

For reconstructing the J-coupling matrix we made use of the fact that the J-coupling is equal to the precession velocity. To measure the isolated mutual coupling between two distinct spins, we decoupled the third spin from the remaining system. This can be achieved using the selective recoupling method (*41*). Here, a single SE $\pi$ pulse is addressed to the spin that shall be decoupled right in the middle of the conditional evolution time. Hence, the other spins' evolution is effectively canceled as a result of this individual spin. Like before, one of the coupled spins was prepared in a superposition state, whereas the other was prepared in an energy eigenstate. From the spin precession the coupling among these two spins was deduced.

### Pulse sequence to change between couplings

The pulse sequence that transfers the spin state from one base to another is based on single-ion control with MW pulses. In the experiments we showed the transfer from the magnetic sensitive base $|^2S_{1/2}, F=0\rangle \leftrightarrow |^2S_{1/2}, F=1, m_F=-1\rangle$ (magnetic $\sigma^-$ transition) to the (first order) insensitive base $|^2S_{1/2}, F=0\rangle \leftrightarrow |^2S_{1/2}, F=1, m_F=0\rangle$ formed by the magnetic $\pi$ transition. The particular sequence for this transfer of spin $n$ consists of three MW pulses and reads

$$\pi_\pi \pi_\sigma^{(n)} \pi_\pi, \quad (10)$$

where $\pi_\pi$ denotes a $\pi$ pulse on the magnetic $\pi$ transition and $\pi_\sigma^{(n)}$ denotes a $\pi$ pulse on qubit $n$'s $\sigma$ transition. The phase of all pulses is $\varphi = 0$.

When using this sequence to decouple a spin by transferring it from the $\sigma$ to the $\pi$ base, it is possible that a small detuning of the pulse that prepares the superposition causes this spin to precess during the conditional evolution time. To suppress this spurious effect, an SE $\pi$ pulse is included into the experimental sequence. After half of the desired evolution time the isolated spin is recoded back to the sensitive base and an SE pulse of phase $\varphi = \pi/2$ on the sensitive transition is applied before it is recoded back to the insensitive base. Right after the complete evolution time the spin is again recoded back to the sensitive base. Now a possible Ramsey experiment can probe the state of this spin $n$. In summary, the spin is recoded four times during the experimental sequence to decouple it.

For the case where more qubits (if not all) are recoded, the sequence in Eq. (10) is changed to

$$\pi_\pi \pi_\sigma^{(1)} \pi_\sigma^{(2)} \pi_\sigma^{(3)} \pi_\pi . \qquad (11)$$

**Dynamical decoupling**

The dynamics within the three-spin system are driven by J-coupling which is typically about $2\pi \times 40$ Hz for next-neighbor spins using a static magnetic gradient of 19 T/m at a center-of-mass frequency, $\nu_1$, of the ions' motion of about $2\pi \times 130$ kHz. Because the single spins' coherences decay in the present experiment on the time scale of 200(100) μs these dynamics are slower than the dephasing. To stabilize the dynamics and to enhance the coherence time we applied DD pulses (*40*) during the conditional evolution times. To keep the desired conditional dynamics unaffected by the DD pulses they need to be applied to all qubits simultaneously. Because the MW source used at present generates a single frequency at a certain time and the qubits' addressing frequencies differ, the DD pulses were applied to the qubits one after each other.

The timing of the DD pulse sequence used to measure the J-coupling matrix corresponds to a Carr-Purcell-Meiboom-Gill (CPMG) sequence with 20 pulses and the time evolution reads

$$[\tau \, \pi_\sigma^{(1)} \pi_\sigma^{(2)} \pi_\sigma^{(3)} \, \tau]^{20} , \qquad (12)$$

where $\tau$ denotes a waiting time, $\pi_\sigma^{(n)}$ denotes a $\pi$ pulse on qubit $n$'s magnetic $\sigma$-transition and the exponent denotes a repetition of the evolution in the brackets. The phase relation of the $\pi$ pulses corresponds to a Knill DD (KDD) sequence (*63*).

Although more pulses may further enhance the coherence time and hence the fringe contrast, after a certain duration the pulse sequence becomes more and more susceptible to instrumental imperfections (*40*). In our setup the pulse errors are consequences of a slow drift of the spins' addressing frequencies. With 84 pulses for example it was possible to protect conditional spin dynamics for 11 ms (*29,40*). Thus, more pulses may enhance the contrast after 4 ms, but at the same time reduce it for shorter evolution times. In general there exists an optimal number of pulses for each evolution time. Twenty DD pulses turned out to be a good compromise for evolution times between 1 and 5 ms as used here to measure the conditional phase shift.

The realization of the QFT took 8.6 ms; hence, more DD pulses were necessary to stabilize the dynamics during this period. During the QFT a total of 60 DD pulses were addressed to each qubit. Twenty pulses were applied during the first conditional evolution time $T_1$ and 40 pulses were applied during $T_3$. These pulse numbers turned out to be an optimal choice, and more pulses did not improve the performance in terms of the fidelity. The reason is that for more pulses the sequence becomes more susceptible to instrumental errors which counteract the improvements due to a longer coherence time.

**Derivation of the QFT sequence**

We base our implementation on the more general approach presented by Ivanov *et al.* (*52*). The $N=3$ Fourier gate can be obtained with the circuit $\Phi_F H_3 e^{i\pi\left(\frac{1}{8}\sigma_z^{(2)}\sigma_z^{(3)}\right)} H_2 e^{i\pi\left(\frac{1}{8}\sigma_z^{(1)}\sigma_z^{(2)} + \frac{1}{16}\sigma_z^{(1)}\sigma_z^{(3)}\right)} H_1 \Phi_I$, where $H_k$ is the Hadamard gate, applied to ion $k$, and

$$\Phi_{\mathrm{I}} = e^{-i\frac{\pi}{4}\sum_{k=1}^{N}(1-2^{-k+1})\sigma_z^{(k)}}, \qquad (13)$$

$$\Phi_{\mathrm{F}} = e^{-i\frac{\pi}{4}\sum_{k=1}^{N}(1-2^{k-N})\sigma_z^{(k)}}, \qquad (14)$$

are the phase gates, applied to all $N = 3$ ions in the beginning and at the end of the circuit. Note that the phase gates represent a shift of the phase of the driving field, rather than a physical modification of the qubit. Hence, they are not associated with infidelity. The circuit can be written in a more explicit form (up to a global phase of $-\pi/2$)

$$\Phi_{\mathrm{F}} R_3\left(\tfrac{\pi}{2},-\tfrac{\pi}{2}\right) R_3(\pi) e^{i\pi\left(\frac{1}{8}\sigma_z^{(2)}\sigma_z^{(3)}\right)} R_2\left(\tfrac{\pi}{2},-\tfrac{\pi}{2}\right)$$
$$R_2(\pi) e^{i\pi\left(\frac{1}{8}\sigma_z^{(1)}\sigma_z^{(2)}+\frac{1}{16}\sigma_z^{(1)}\sigma_z^{(3)}\right)} R_1\left(\tfrac{\pi}{2},-\tfrac{\pi}{2}\right) R_1(\pi) \Phi_{\mathrm{I}}, \qquad (15)$$

where we have used the Hadamard gate decomposition $H = iR(\pi/2,-\pi/2)R(\pi)$. Below we will implement this circuit with our gate set.

An operation that requires particular attention is the implementation of the term $e^{i\pi\left(\frac{1}{8}\sigma_z^{(1)}\sigma_z^{(2)}+\frac{1}{16}\sigma_z^{(1)}\sigma_z^{(3)}\right)}$. During this process, no net evolution occurs between qubits 2 and 3, whereas a conditional evolution occurs between qubits 1 and 2 and between qubits 1 and 3. One possible realization is the consecutive realization of the coupling topologies where only qubits 1 and 2 or qubit 1 and 3 are coupled. In particular one could first isolate qubit 3 for a certain time interval and then isolate qubit 2 for an additional time interval. However, such a realization would be time consuming and a parallel decomposition seems advantageous. For the implementation of this operation we note that if we surround the gate $U(T)$ (which is based on all the couplings among the system being present) with two $\pi$ pulses on one of the qubits, we effectively change the sign of the coupling of the same ion with the other ions. If we apply the $\pi$-pulses to ion 3, we change all couplings $J_{jk}$, where index 3 is present:

$$R_3(\pi) e^{\frac{iT}{2}\left(J_{12}\sigma_z^{(1)}\sigma_z^{(2)}+J_{13}\sigma_z^{(1)}\sigma_z^{(3)}+J_{23}\sigma_z^{(2)}\sigma_z^{(3)}\right)}$$
$$\times R_3(\pi) = e^{\frac{iT}{2}\left(J_{12}\sigma_z^{(1)}\sigma_z^{(2)}-J_{13}\sigma_z^{(1)}\sigma_z^{(3)}-J_{23}\sigma_z^{(2)}\sigma_z^{(3)}\right)}. \qquad (16)$$

Thus the sequence $U(T_2)R_3(\pi)U(T_1)R_3(\pi)$ yields

$$e^{\frac{i}{2}\left((T_2+T_1)J_{12}\sigma_z^{(1)}\sigma_z^{(2)}+(T_2-T_1)J_{13}\sigma_z^{(1)}\sigma_z^{(3)}+(T_2-T_1)J_{23}\sigma_z^{(2)}\sigma_z^{(3)}\right)}. \qquad (17)$$

Now if we choose

$$T_1 = \frac{\pi}{8}\left(\frac{1}{J_{12}}+\frac{1}{2J_{13}}\right), \qquad (18)$$

$$T_2 = \frac{\pi}{8}\left(\frac{1}{J_{12}}-\frac{1}{2J_{13}}\right), \qquad (19)$$

we get the term $e^{i\pi\left(\frac{1}{8}\sigma_z^{(1)}\sigma_z^{(2)}+\frac{1}{16}\sigma_z^{(1)}\sigma_z^{(3)}\right)}$ and a residual factor. This factor can be eliminated with properly selected rotations, determined from the expansion

$$e^{i\phi\sigma_k\sigma_p} = \cos\phi\,\mathbf{1} + i\sin\phi\,\sigma_k\sigma_p, \qquad (20)$$

where $k,p = x,y,z$.

Thus we rearrange the above sequence, which now acquires the form

$$R_2\left(A_2, \tfrac{3\pi}{4}\right) R_3\left(\tfrac{\pi}{2}, -\tfrac{\pi}{2}\right) U_{23}(T_3)$$
$$R_1\left(\pi, \tfrac{3\pi}{16}\right) R_2\left(A_1, \tfrac{3\pi}{4}\right) R_3\left(\pi, -\tfrac{3\pi}{16}\right)$$
$$U(T_2) R_3(\pi) U(T_1) R_1(\pi) R_2(\pi) R_3(\pi) H_1, \quad (21)$$

where the duration $T_3$ and the pulse areas $A_1$ and $A_2$ are obtained from the following set of equations

$$\frac{1}{\sqrt{2}} e^{i\frac{\pi}{16}(\alpha+2)} e^{iJ_{23}T_3/2} - \sin\frac{A_1}{2}\sin\frac{A_2}{2} e^{iJ_{23}T_3}$$
$$+ \cos\frac{A_1}{2}\cos\frac{A_2}{2} = 0, \quad (22)$$

$$\frac{1}{\sqrt{2}} e^{i\frac{\pi}{16}(\alpha-2)} e^{iJ_{23}T_3/2} - \sin\frac{A_1}{2}\cos\frac{A_2}{2} e^{iJ_{23}T_3}$$
$$- \cos\frac{A_1}{2}\sin\frac{A_2}{2} = 0. \quad (23)$$

Here $\alpha = J_{23}/J_{13}$. Note that the sequence in Eq. (21) is generic in the sense that it is valid for *any* spin-spin coupling $J$. This sequence was further optimized as described in the main text.

**State fidelities**

To validate the QFT, we implemented the appropriate pulse sequence after initializing the three-qubit system in one of the computational basis states or in some exemplary superposition states. In the next step we reconstructed the fidelity $F = \langle\psi|\rho|\psi\rangle$ of the three-qubit state $\rho$ after the QFT given the ideal result $|\psi\rangle$. Further analysis takes into account the single-qubit fidelities in each qubit's subspace, $F_1, F_2$, and $F_3$, and the full three-qubit state fidelity $F$.

The QFT transfers the input states investigated here to fully separable product states $|\psi\rangle = |\psi_1\rangle|\psi_2\rangle|\psi_3\rangle$. Under the assumption of a fully separable state after the QFT, $\rho = \rho_1 \times \rho_2 \times \rho_3$ we derived the single-qubit state fidelities $F_n = \langle\psi_n|\rho_n|\psi_n\rangle$ from Ramsey fringes (phase and amplitude).

The three-qubit state fidelity is measured without any assumption on $\rho$. The fidelity $F = \langle\psi|\rho|\psi\rangle$ can be written as $F = \langle 000|V^\dagger \rho V|000\rangle$, where $V$ denotes local rotations that connect the ideal outcome $|\psi\rangle$ with the ground state $|000\rangle$. The procedure is therefore to apply appropriate single-qubit rotations $V$ at the end of the QFT just before the projective measurement takes place. The probability of detecting the state $|000\rangle$ then yields the fidelity $F$. The results are summarized in Table 1.

**Table 1: State fidelities after application of the QFT.** The single-qubit state fidelities $F_1, F_2$ and $F_3$ are estimated from Ramsey fringes and excitation probabilities after applying the QFT. From correlation measurements the three-qubit state fidelities $F$ are deduced.

| Initial state | $F_1$ | $F_2$ | $F_3$ | $F_1 \times F_2 \times F_3$ | $F$ |
|---|---|---|---|---|---|
| $\|000\rangle$ | 0.74(5) | 0.88(5) | 0.82(5) | 0.53(6) | 0.59(2) |
| $\|001\rangle$ | 0.77(4) | 0.81(2) | 0.86(4) | 0.53(4) | 0.54(2) |
| $\|010\rangle$ | 0.76(7) | 0.86(4) | 0.84(4) | 0.55(4) | 0.55(2) |
| $\|011\rangle$ | 0.77(5) | 0.84(2) | 0.83(4) | 0.54(5) | 0.54(2) |
| $\|100\rangle$ | 0.74(5) | 0.89(5) | 0.88(5) | 0.58(6) | 0.59(2) |
| $\|101\rangle$ | 0.73(4) | 0.85(2) | 0.90(4) | 0.56(4) | 0.66(2) |
| $\|110\rangle$ | 0.70(4) | 0.78(4) | 0.90(3) | 0.49(4) | 0.57(2) |
| $\|111\rangle$ | 0.68(4) | 0.79(3) | 0.89(4) | 0.48(4) | 0.63(2) |
| $\|+00\rangle$ | 0.84(2) | 0.78(5) | 0.86(5) | 0.56(5) | 0.65(2) |
| $\|+01\rangle$ | 0.83(2) | 0.84(3) | 0.81(4) | 0.56(4) | - |
| $\|+10\rangle$ | 0.81(2) | 0.77(4) | 0.76(4) | 0.47(3) | - |
| $\|+11\rangle$ | 0.82(2) | 0.81(2) | 0.80(4) | 0.54(3) | - |
| $\|++0\rangle$ | 0.79(2) | 0.74(2) | 0.88(4) | 0.51(3) | - |
| $\|++1\rangle$ | 0.84(2) | 0.73(2) | 0.75(4) | 0.46(3) | - |
| $\|+++\rangle$ | 0.86(2) | 0.79(2) | 0.77(2) | 0.52(3) | 0.54(2) |

We point out that we have carefully selected the input states in Table 1 in the following sense: With the first eight measurements we made sure that the QFT is applied correctly to all eight computational basis states. Thus, we verified that each column vector in the measured QFT matrix is correct, possibly up to a column-specific global phase, which may vary from column to column.

Next we verified that these column-specific phases are equal, so that the correct QFT matrix is obtained up to a global phase. For this purpose we carried out seven additional measurements on superpositions. The first four measurements indicate that column 1 is in phase with column 5, column 2 is in phase with column 6, column 3 is in phase with column 7, and column 4 is in phase with column 8. The last three measurements verify that column 1 is in phase with column 3, column 2 is in phase with column 4, and column 1 is in phase with column 2. With the described measurements we validated the QFT matrix up to a global phase.

For the computational basis states the average state fidelity is $\langle F \rangle = 0.58(5)$. As one can also see, the results do not significantly differ for the investigated computational basis states and the exemplary superposition states. One may further notice that the product $F_1 \times F_2 \times F_3$ that is equal the three-qubit state fidelity under the assumption of fully separable final states and the measured three-qubit state fidelities $F$ are in agreement.

**Technical imperfections in the setup**

Experimental imperfections in the setup are sources of error. The most dominant source at present is the magnetic field noise that causes dephasing of the qubits. During the conditional evolution time, magnetic field noise is counteracted by DD pulses. However, these pulses are not perfect due to slow drifts of the qubits' addressing frequencies (*40*) and can cause a seemingly random behavior of the qubits, when many DD pulses are applied. Another instrumental error is the nonperfect single-shot detection efficiency. This causes a mistagging of the results and, hence, effectively a seemingly random outcome as well.

These different error sources have consequences for the performance of the QFT that differ for different input states of the algorithm. Qubits that are in a superposition state during the sequence are affected by dephasing, whereas a qubit in an energy eigenstate is not. Also for the histograms from which the periods are deduced (compare Fig. 4 in the main text) one may notice a very good performance for the input state $|111\rangle$. The reason is that the expected ideal histogram of this output state is the same as that for a totally dephased mixture of states. Therefore, the results are almost not affected by the present technical imperfections.

To date, there is no magnetic shielding or active stabilization of the magnetic field at the location of the ions. Adding this is expected to substantially improve coherence times and fidelities. In addition, the single-shot readout fidelity of the experiments reported here is 0.96 for a bright or a dark state and will be improved either by more advanced analysis (*64*) or by changes of the optical detection setup (*65*).

**Simulations**

In independent experiments we experimentally quantified the exponential decay of single-qubit coherences during the conditional evolution times. Hence, we can simulate the dephasing that leads to the partially dephased three-qubit density matrix $\rho_{\text{dephased}}$. In this matrix the single-qubit coherences are partially decayed and the particular decay depends on the duration during which a single qubit was in a superposition state. The experimentally determined decay constant is 0.0625/ms.

Furthermore, we deduced from separate experiments the effect of a seemingly chaotic behavior due to imperfect DD pulses and a limited single-shot readout fidelity. These two effects cause a change between the qubit states $|0\rangle$ and $|1\rangle$ and can be treated by mixing the three-qubit density matrix with white noise: $\tilde{\rho}_{\text{simulated}} = \zeta \mathbf{1} + (1-\zeta)\rho_{\text{dephased}}$.

The resulting fidelity affected by dephasing of the qubits and by white noise reads
$$\tilde{F} = \zeta/8 + (1-\zeta)F_{\text{dephased}}. \qquad (24)$$
Here, $F_{\text{dephased}}$ describes the fidelity due to pure dephasing without any further instrumental errors and the phenomenological parameter $\zeta = 0.25$ describes the amount of white noise in the final state. These phenomenological parameters (the decay constant and $\zeta$) allow us to simulate the time evolution of the pulse sequence that realizes the QFT in a way that includes the experimental imperfections.

We compared the simulated results to the recorded data for the experiments in which the periods of quantum states were computed by our quantum computer prototype. In detail, we computed the SSO between the measured probabilities $p_i$ and the simulated outcomes $s_i$ as $S^{sim}(\{p_i\},\{s_i\}) = (\sum_i \sqrt{p_i \times s_i})^2$. For all data sets we obtained an SSO close to unity and the average of all data sets was 0.98(-3+2). Therefore, we conclude that the limitations of the experimental results are dominated by the experimental imperfections that are part of our model.

**Error budget and perspectives**

On the basis of our measurements and simulations of the experiment, we now further discuss how the infidelities due to the different error sources contribute to the average fidelity of $\tilde{F} = 0.58(5)$. Using Eq. (24) and setting $\zeta = 0$, we conclude that if the system were subjected to residual dephasing only, the fidelity would be limited to $F_{\text{dephased}} = 0.73$. White noise only (this is, considering only errors in detection and DD pulses by setting $F_{\text{dephased}} = 1$ in Eq. (24) would limit the resulting fidelity in Eq. (24) to $\tilde{F} = 1 - 0.22 = 0.78$ (for $\zeta = 0.25$), where $0.12 = 1 - 0.96^3$ is the contribution to the infidelity from the limited single-shot detection fidelity and the remaining infidelity (0.10) is caused by errors in the DD sequences.

This error budget allows us to predict the possible performance in an improved experimental setup. In the present setup the magnet gradient experienced by the ions is 19 T/m. If the gradient is enhanced to 150 T/m, as was recently reported by Lekitsch *et al.* (*60*), the coupling strengths, being proportional to the gradient squared, will be about a factor of 60 higher. Therefore, such a gradient will speed up the algorithm by the same factor resulting in a total duration of only about 140 μs.

In this situation there would be no need for sequences of DD pulses because a single SE pulse already enhances the coherence time to a few milliseconds in our setup. Hence the error due to imperfect DD pulses will be eliminated. For discussing the fault tolerance of a quantum gate, state preparation and detection errors can be usually neglected. Therefore, the gate fidelity will only be limited by errors of the few remaining pulses that realize the QFT. Because MW-driven single-qubit gates can be realized with high fidelity (*31*), a gate fidelity of the QFT in the range of 0.99 is realistic in a system that features a gradient of about 150 T/m. Measures to increase the qubit coherence time (shielding and active stabilization of magnetic fields) would further increase this fidelity.

**Acknowledgments**
**Funding:** We acknowledge funding from the Deutsche Forschungsgemeinschaft, the Bundesministerium für Bildung und Forschung (FK 16KIS0128), and the European Community's Seventh Framework Programme (FP7/2007-2013) under grant agreement no. 270843 (iQIT).
**Author contributions:** C.W. proposed the project and conceived the idea. C.P. and T.S. carried out the experimental study. S.S.I. proposed and developed the realization of the QFT. S.W. assisted in data analysis and interpretation of the results. C.P. and C.W. wrote the paper. All authors analyzed the data and commented on the manuscript.
**Competing interests:** The authors declare that they have no competing interests. **Data and materials availability:** All data needed to evaluate the conclusions in the paper are present in the paper. Additional data related to this paper may be requested from the authors.